\documentstyle[12pt]{article}
\begin{document}
\renewcommand{\thefootnote}{\fnsymbol{footnote}}
\begin{titlepage}
\begin{flushright}
KEK-TH-429 \\
KEK-CP-26 \\
KEK preprint 94-201 \\
TIT/HEP-286 \\
March 1995 
\end{flushright}
\vspace*{10mm}
\Large
\begin{center}
Complex Structures \\
 Defined on \\
Dynamically Triangulated Surfaces
\end{center}
\vspace{10mm}
\large
\begin{center}
H.Kawai, \footnote[2]{E-mail address: kawaih@theory.kek.jp}
N.Tsuda, \footnote[3]{E-mail address: ntsuda@phys.titech.ac.jp\\
Address till March 31st:\\
Department of Physics, Tokyo Institute of Technology,
Oh-okayama, Megro, Tokyo 152, Japan}
and 
  T.Yukawa \footnote[4]{E-mail address: yukawa@theory.kek.jp}
\end{center}
\begin{center}
National Laboratory for High Energy Physics (KEK) \\
       Tsukuba, Ibaraki 305 , Japan
\end{center}
\vspace{1cm}
\parindent 5cm
\begin{abstract}
A method to define the complex structure and separate the conformal mode
is proposed for a surface constructed by two-dimensional dynamical 
triangulation.
Applications are made for surfaces coupled to matter fields such as 
$n$ scalar fields ($n = 0,1$ and $4$) and $m$ Ising spins ($m = 1$ and $3$).
We observe a well-defined complex structure for cases when the matter
central charges are less than and equal to one, while it becomes unstable 
beyond $c = 1$. 
This can be regarded as the transition expected in analytic theories.
\end{abstract}
\end{titlepage}
  
\section{Introduction}
There has been, over the last few years, remarkable progress in the 
quantum theory of two-dimensional(2-d) gravity coupled to matter fields, 
such as scalar fields or Ising spins, connected to string theoretic models.
Two distinct analytic approaches for quantizing 2-d gravity have been 
established, these being the discretized\cite{matrix} and continuous
\cite{Liouville} theories.
The discretized approach, implemented by the matrix model technique, 
exhibits behavior found in the continuous approach, given by Liouville 
field theory, in a continuum limit.
There thus seems to exist strong evidence for the equivalence of the two 
theories. 
However we feel that the mutual relationship is not yet fully understood.

Recently numerical methods, such as that of dynamical triangulation\cite{dt} 
and Regge calculus\cite{Regge}, have drawn much attention as alternative 
approaches to studying non-perturbative effects, being also capable of handling
those cases where analytical theories cannot yet produce meaningful results. 
In the dynamical triangulation method calculations of the partition function 
are performed by replacing the path integral over the metric to a sum over 
possible triangulations.
Dual graphs of surfaces created by this method have direct correspondence 
to Feynman diagrams of the $\phi^3$ matrix model.
While the relation between dynamical triangulation and matrix model is 
rather transparent, its relation to Liouville field theory has not been as 
evident.
However, intuitively, we expect that the manifold
treated in the continuous theory can be approximated by the
triangulated lattice when the number of triangles are large enough.   

On the other hand we still do not understand the nature of the transition 
at $c = 1$, a barrier which is evident in analytical theories.
In spite of serious numerical efforts\cite{AmbjDurh} 
all attempts have failed so far to account for the transition, at least 
by attempting to measure the string susceptibility of 
simulated surfaces.
Of course the string susceptibility, $\gamma_{string}$, is defined 
from the asymptotic behavior of the partition function as
$Z[N] \approx N^{\gamma_{string}-3} e^{\lambda N}$, 
and it is obviously not an easy task to extract $logN$ 
contributions from dominant $\lambda N$ backgrounds 
unless we have good estimates of the finite size correction. 

It follows that it should be much easier to observe this transition in 
local scale invariant quantities, rather than in the original metric which
is comparatively complicated, reflecting the fractal nature of the surface.
In this letter we propose a method by which we can define the complex 
structure of a surface generated by the dynamical triangulation method
and separate the conformal mode.
Thus this enables us to observe the transition by examining the
possibility of defining a smooth background metric. 

\section{Determination of the complex structure }
A fluctuating metric $g_{\mu\nu}$ defined on a two dimensional continuous 
surface can be decomposed into its complex structure(moduli $\tau$) 
and a conformal mode $\phi(z)$ :
\begin{equation}
g_{\mu\nu} = \hat g _{\mu\nu}(\tau,z)e^{\phi(z)}.
\end{equation}
In the lattice formulation of quantum gravity one expects
that surfaces generated by the dynamical triangulation method
will tend to continuous surfaces in the limit of large numbers of triangles.
However, such surfaces are known to be fractal\cite{fractal1,fractal2}, 
and it is not at all trivial that the lattice regularization
simulates the continuous theory for a small lattice constant. 
It follows that one of the indispensable problems in the discretized method 
we are employing is to make sure that we indeed possess a universal continuum 
limit,{\it i.e.} a unique complex structure and conformal mode.

Our method stems from attempts to derive dual amplitudes
from planar Feynman diagrams in the large order limit\cite{fishnet_analog}, 
by making use of the electrical circuit analogy\cite{circuit}.
The basic assumption of these methods is that the network corresponding to 
a fine planar Feynman diagram can be regarded as a uniform homogeneous 
conducting sheet with a constant resistivity.
Once this is accepted it is straightforward to derive dual amplitudes
by evaluating the heat generated by the sheet.
Unfortunately, there exists no realistic field theory in which 
this assumption is literally satisfied so that dual amplitudes
are given as the result.
For non-critical strings, however, we know matrix models with $c \le 1$ 
reproduce results of the continuum theory ($i.e.$ the Liouville theory).
Therefore, we expect the basic assumption is satisfied for these models.
In other words random surfaces corresponding to 2-d quantum gravity coupled 
to $c \le 1$ matter fields should behave as uniform homogeneous media 
with constant resistivities if we regard them as electrical networks.

In order to check the idea numerically it is necessary to measure how the 
network conducts current, and compare it with the corresponding continuous 
medium.
In the following we first give an algorithm to measure the resistivity from 
current and voltage distributions.
As we shall see, the invariance of resistivity under local scale 
transformations plays an essential role in the algorithm, and because 
of this property we can extract information which is independent of the 
conformal mode such as moduli.
We then apply the method to the dynamically triangulated surfaces.
Since local scale invariance is not fully realized on discretized surfaces, 
the resistivity thus obtained may have certain fluctuations.
In fact, the value varies from one sample surface to another, 
and with the position of electrodes for the measurements.
Such ambiguities will vanish, however, in the continuum limit, 
if it exists at all. 
We shall see that this is indeed the case for matrix models with $c \le 1$.
   
Let us first consider a two dimensional conducting medium with conductivity 
tensor given by $\sigma^{\mu \nu}$.
The Joule heat $Q$ generated by a potential distribution $V$ on the surface,
\begin{equation}
Q=\int d^2x \sigma^{\mu \nu} \partial_{\mu}V \partial_{\nu}V ,
\end{equation}
leads to an equation for $V$ by requiring, $\delta Q=0$;
\begin{equation}
\partial_{\mu}\sigma^{\mu \nu} \partial_{\nu}V = 0.
\label{eq:eqofmotion}
\end{equation}
When we define the current density by
\begin{equation}
j^{\mu}= \sigma^{\mu \nu}\partial_{\nu}V ,
\label{eq:current}
\end{equation}
eq.(\ref{eq:eqofmotion}) is just the equation of  continuity.
By identifying $\partial_{\nu}V$ to be the electric field $E_{\nu}$
eq.(\ref{eq:current}) becomes, of course, Ohm's law, 
$j^{\mu}=\sigma^{\mu \nu}E_{\nu}$, or as it is often expressed 
$E_{\mu}=\rho_{\mu \nu}j^{\nu}$ with the resistivity tensor, 
$\rho_{\mu \nu}$, defined by
\begin{equation}
\sigma^{\mu \lambda} \rho_{\lambda \nu}=\delta^\mu_\nu .
\end{equation}

For a uniform homogeneous medium whose surface is specified by the metric 
tensor $g_{\mu \nu}$ the resistivity tensor is written as
\begin{equation}
\rho_{\mu \nu}= r {1 \over \sqrt {g}} g_{\mu \nu} ,
\end{equation}
with $r$ being the resistivity constant.  
An important property of the resistivity tensor in two dimensions is the 
invariance under local scale transformations,
$ g_{\mu \nu} \mapsto  g_{\mu \nu} e^{-\sigma} $.
The significance of this invariance property is seen by considering the 
resistance $R$ of a small rectangular section of the conducting sheet with 
length $a$ and width $b$:  
\begin{equation}
R={a \over b}r ,
\end{equation}
which is invariant under the scale change of 
$a\mapsto a\xi$ and $b\mapsto b\xi$.
This property enables us to determine the complex structure of two 
dimensional curved surfaces through measurements of the resistivity.

In the case of spherical topology we can regard the surface
as an infinite flat sheet. 
Then the potential at a point with the (complex) coordinate, 
$z(=x+{\it i}y)$, 
with a source of current $I$ placed at $z_3$, and a sink 
of the current at $z_4$ is  
written as 
\begin{equation}
V_{34}(z)=-{Ir \over 2\pi} \ln  |{z-z_3 \over z-z_4}| + {\rm Const}.,
\end{equation}
where the subscript of $V$ indicate points of the source 
and the sink.
In order to avoid the ambiguity arising from Const. in the above equation 
we measure the potential drop between points $1$ and $2$ 
with coordinates $z_1$ and $z_2$ (Fig.1),
\begin{equation}
V_{34}(z_1)-V_{34}(z_2)=-{Ir \over 2\pi} \ln |[z_1,z_2;z_3,z_4]| ,
\label{eq:potential}
\end{equation}
where the ratio of four points written as
$$[z_1,z_2;z_3,z_4]=
{z_1-z_3 \over z_1-z_4}{z_2-z_4 \over z_2-z_3}
$$
is known as the anharmonic ratio\cite{EDM}.

The anharmonic ratio has several interesting properties.
For example, its invariance under 
the projective transformation,
\begin{equation}
z \mapsto {az+b \over cz+d}   ,
\end{equation}
with $ad-bc=1$ allows us to fix three coordinates among the $\{z_i\}$ to 
any desired values without changing the potential drop eq.(\ref{eq:potential})
by appropriately choosing three complex parameters among $a,b,c$, and $d$. 
For example, we can fix three coordinates
as $z_2=1$, $z_3=0$, and $z_4=\infty$. 
Then the potential drop eq.(\ref{eq:potential}) for $I = 1$A is written as
\begin{equation}
V_{34}^{12}=-{r \over 2\pi} \ln | z_1| ,
\label{eq:potd1}
\end{equation}
where $V_{34}^{12}$ stands for the potential drop
between points 1 and 2 in the presence of a source and a sink
of current $I=1$A at points 3 and 4, respectively.

In addition, by permuting the four points the anharmonic ratios 
are known to take values such as
\begin{eqnarray}
[z_1,z_2;z_3,z_4] & = & \lambda, \cr
[z_1,z_3;z_2,z_4] & = & 1-\lambda, \cr
[z_1,z_2;z_4,z_3] & = & 1/\lambda, \cr
[z_1,z_4;z_2,z_3] & = & 1-1/\lambda.
\end{eqnarray}
For other possible combinations they are given by
ratios of these complex numbers.

Using the second of the above equations we get 
\begin{equation}
V_{24}^{13}=-{r \over 2\pi} \ln | 1-z_1| .
\label{eq:potd2}
\end{equation}
All the other possible permutations of $\{z_i\}$ only give 
the potential drops
obtained by linear combinations of eq.(\ref{eq:potd1}) and
(\ref{eq:potd2}). 
Therefore, using this set we can determine $z_1$ within the uncertainty 
of its complex conjugate, if we know the resistivity constant $r$ from 
other means. 
There may be several methods to determine $r$. 
We describe here one of the methods\footnote{We will refer to this method 
as the five point method in later discussion.} which we have employed 
in our numerical simulations.
If we introduce an additional point for the measurement(Fig.2),
we get three more relations in addition to  eq.(\ref{eq:potd1}) and
(\ref{eq:potd2})
which together are sufficient to specify five unknowns,
namely two points in the complex plane and $r$. 
Denoting the coordinate added as $z_5$
the three extra equations are given by
\begin{eqnarray}
V_{34}^{52}&=&-{r \over 2\pi} \ln | z_5| ,\\
V_{24}^{53}&=&-{r \over 2\pi} \ln | 1-z_5| , \\
V_{14}^{52}&=&-{r \over 2\pi} \ln |[z_5,z_2;z_1,z_4] |\nonumber\\
          &=&-{r \over 2\pi} \ln | {z_5-z_1 \over 1-z_1}|.   
\end{eqnarray}

Now we apply this algorithm to a random surface generated by the 
dynamical triangulation method. 
The dual graph of a surface consisting of $N$ triangles
is regarded as a trivalent network, where we fix the resistance
of the link connecting two neighbouring vertices to be $1\Omega$.
What we want to do is to examine whether such a network
behaves like a continuous medium in the limit of $N \rightarrow \infty$
or not.
For the measurement we pick five vertices in the dual graph and perform 
the five point method as explained above.
We note that this sometimes produces two solutions for $r$ due to the 
uncertainty explained below eq.(\ref{eq:potd2}). 
This, however, is not a great problem, since the probability of such cases 
is about 10 to 20 per cent, so we can simply throw them away without 
adversely affecting the statistics.

The practical method we employ for the determination of potential drops
is as follows;
we pick two vertices at $P_{in}$ and $P_{out}$ for the source and the sink 
of current with unit intensity($1A$).
By writing the potential of the vertex at $P$ as $V(P)$ current conservation 
reads
\begin{equation}
{V(P)-V(P_a) \over 1}+{V(P)-V(P_b) \over 1}+{V(P)-V(P_c) \over 1}=0,
\label{eq:currentcons}
\end{equation}
expressing the zero net flow to the three
neighboring vertices $P_a,P_b$ and $P_c$, which gives
$$
V(P)={1 \over 3}\{V(P_a)+V(P_b)+V(P_c)\} ,
$$
except at two vertex points where the current enters or exits,
$$
V(P_{in})={1 \over 3}\{V(P_a)+V(P_b)+V(P_c)+1\},
$$
and
$$
V(P_{out})={1 \over 3}\{V(P_a)+V(P_b)+V(P_c)-1\},
$$
respectively.
It is straightforward to obtain potentials at all the vertex points
on the surface numerically by the Jacobi iteration method
\footnote{In practice we employ the successive over-relaxation(SOR) 
method for rapid convergence.}. 

\section {Numerical results and discussions}
Before presenting our numerical results let us sketch our simulation.
Random surfaces are generated by the dynamical triangulation method
with fixed topology($S^2$) and a fixed number of triangles 
(the micro-canonical simulation) prohibiting tadpole and self-energy 
diagrams.
For matter fields coupled to gravity we put $n$ scalar fields with 
$n = 0,1$ and $4$ and/or $m$ Ising spins $m = 1$ and $3$ on the triangle.
For thermalization of Ising spins we employ the Wolf algorithm\cite{Wolff} 
to avoid severe critical slowing down.

The distribution of the resistivity constant is determined in the following 
manner;
we first pick-up 40 to 50 independent configurations from the ensemble
\footnote{Here we consider dual graphs of triangulated surfaces.}.
For each configuration we pick five vertices randomly to 
perform the five point method to determine the resistivity.
We have repeated this procedure 50 times for each configuration.
Thus a graph of the distribution of $r$ consists in total of about 2,000 
to 2,500 data values.

Let us first take a look at the measurement for pure gravity(Fig.3).
The distributions of $r$ for three different lattice sizes
show distinct peaks at about $2.6$, and the {\it peaks get narrower as the 
size grows}.
The value $r \approx 2.6$ should be compared to $\sqrt {3}$ of
the flat network($i.e.$ 6 triangles around each apex).
An increase in $r$ is understood to be a reflection of the fractal nature 
of the surface.
As the number of triangles gets larger, finite size effects due to
the discreteness of the surface diminish and eventually the peak
grows infinitely. 
This is what we have expected as the continuum limit of a network of 
resistors.
We find similar tendency in the one Ising($c={1 \over 2}$) and the 
one scalar ($c=1$) (Fig.$4$ $(a)$) cases. 

While random surfaces coupled to matter with central charge less than or 
equal to one  seem to approach to a uniform homogeneous medium, simulations 
with matter central charge larger than one shows a different behavior.
In Fig.$4$ $(b)$ the distributions of $r$ in the $c=4$ case is 
compared for the sizes $N=2,000$ and $8,000$.
Here no sharpening of peaks as increasing sizes 
is seen unlike the previous cases, which suggests
that the surface does not approach a smooth continuum limit due to 
instability, which is expected from the complex string susceptibility 
in the Liouville theory.
Existence of the transition at about $c=1$ is also supported in the 
simulations with 3 Ising spins(Fig.5).
This system behaves as a model with $c = {3 \over 2}$ matter at the 
critical point, and $c=0$ off the critical point. 
We observe the peaking phenomena only in the latter case as the size 
increases from $2,000$ to $8,000$.

In conclusion we have observed the so called $c=1$ barrier for two 
dimensional quantum gravity coupled to matter as the transition from a 
well-defined complex structure to an ill-defined one. 
Although our simulation with 8,000 triangles has been enough to observe 
the indication of such a transition, larger size simulations should 
exhibit the phenomena much more clearly.

Next, we discuss how to extract information on the conformal mode.
We have seen, so far, that the resistivity of the dual graph of a dynamically
triangulated surface takes a well-defined value, if the matter central 
charge is less than or equal to one.
Once we have found the value of the resistivity, we can assign a complex 
coordinate to each vertex of the dual graph from  
eq.(\ref{eq:potd1}) and (\ref{eq:potd2}).
Although these two equations alone cannot specify in which half-plane
(upper or lower) it lies, this ambiguity can be easily resolved by the 
five point method.

When we plot the obtained coordinates of all the vertices on a complex plane 
the point density around a point $z$ should be proportional to 
$\sqrt {g(z)}$, because each vertex is supposed to carry the same space-time 
volume.
In the conformal gauge which we have been employing, we have 
$\sqrt {g(z)} = \exp {\phi(z)}$, where $\phi(z)$ is the conformal mode.
The value of the conformal mode is then given by taking the logarithm of the 
point density. 
This argument may be over-simplified in the sense that we did not pay 
attention to the quantum fluctuation of $\phi(z)$ and the renormalization 
of the composite operator $\exp {\phi(z)}$.
What we can say safely is that the point density at $z$ after taking an 
ensemble average should be equal to the following expression in Liouville 
theory,
$$ 
< e^{\alpha\phi(0)}e^{\alpha\phi(1)}
e^{\alpha\phi(\infty)}e^{\alpha\phi(z)} > 
$$
with $\alpha$ being the renormalization factor for the conformal mode.

 Lastly, we briefly discuss a way to define complex structures 
on dynamically triangulated surfaces with higher genus.
For simplicity here we consider the case of a torus, but the generalization 
to higher genus would be straightforward.
In the continuous formulation the period $\tau$ is obtained by the 
following procedure. 
First we introduce harmonic $1$-forms $j_{\mu}dx^{\mu}$ which by 
definition satisfy the divergence and rotation free conditions 
\begin{equation}
\partial_{\mu}j^{\mu} = 0,
\label{eq:div_fre}
\end{equation}
and
\begin{equation}
\partial_{\mu}j_{\nu} - \partial_{\nu}j_{\mu} = 0,
\label{eq:rot_fre}
\end{equation}
where $j^{\mu} = \sqrt{g} g^{\mu \nu} j_{\nu}$.
Since there are two linearly independent solutions, we can impose 
two conditions such as
\begin{equation}
\oint_{a} j_{\mu} dx^{\mu} = 1,
\end{equation}
and
\begin{equation}
\oint_{a} \tilde{j}_{\mu} dx^{\mu} = 0,
\end{equation}
where $\tilde{j}_{\mu}$ is the dual of $j_{\mu}$ given by 
\begin{equation}
\tilde{j}_{\mu} = \epsilon_{\mu \nu} \sqrt{g} g^{\nu \lambda} j_{\lambda},
\end{equation}
and $a$ is a cycle on the torus. 
Then the period $\tau$ is defined by 
\begin{equation}
\tau = \oint_{b} j_{\mu} dx^{\mu} + i\oint_{b} \tilde{j}_{\mu} dx^{\mu},
\end{equation}
where $b$ is a cycle which intersects with $a$ just once.

 This procedure can be easily translated to the case of a triangulated 
surface by identifying $j_{\mu}$ with the current on the dual graph. 
With this identification (\ref{eq:div_fre}) and (\ref{eq:rot_fre}) 
respectively mean current conservation and zero potential drop along 
contractible loops. 
It is easy to show that two linearly independent solutions exist as in 
the continuous case. 
Then the rest is almost trivial, once one recognizes that 
$\oint_{a} j_{\mu} dx^{\mu}$ and $\oint_{a} \tilde{j}_{\mu} dx^{\mu}$
respectively mean the potential drop around $a$ and the total current 
going across $a$.

\begin{center}
{\Large Acknowledgements}
\end{center}
We are grateful to B.E.Hanlon, N.D.Haridass and N.Ishibashi for 
useful discussions and comments.
One of the authors(N.T.) is supported by Research Fellowships of the Japan 
Society for the Promotion of Science for Young Scientists.

\newpage


\begin{thebibliography}{99}
\bibitem{matrix} 
E.Br\'{e}zin and V.Kazakov, 
Phys.Lett. 236B (1990) 144;

M.Douglas and S.Shenker, 
Nucl.Phys. B335 (1990) 635; 

D.Gross and A.Migdal, 
Phys.Rev.Lett. 64 (1990) 127.

\bibitem{Liouville} 
V.G.Knizhnik, A.M.Polyakov and A.B.Zamolodchikov, 
Mod.Phys.Lett A, Vol.3 (1988) 819; 

J.Distler and H.Kawai, 
Nucl.Phys. B321 (1989) 509.

F.David, 
Mod.Phys.Lett. A3 (1988) 1651;

\bibitem{dt} 
F.David, 
Nucl.Phys. B257[FS14] (1985) 45;

V.A.Kazakov, 
Phys.Lett. B150 (1985) 282;

J.Ambj\o rn, B.Durhuus and J.Fr$\ddot{o}$hlich, 
Nucl.Phys.B257[FS14] (1985) 433; Nucl.Phys.B275 [FS17] (1986) 161.


\bibitem{Regge} 
H.W.Hamber,
{\em Critical Phenomena, Random Systems, Gauge Theories, 
Proceedings of the Les Houches Summer School 1984}, 
eds. K.Osterwalder and R.Stora, North-Holland (1986).

\bibitem{AmbjDurh} 
J.Ambj\o rn, B.Durhuus, T.J\'onsson 
and G.Thorleifsson, Nucl.Phys. B398 (1993) 568;

S.M.Catterall, J.B.Kogut and R.L.Renken, 
Phys.Lett. B292 (1992) 277 ; Phys.Rev. D45 (1992) 2957; 

G.Thorleifsson, 
Nucl.Phys. B(Proc. Suppl.) 30 (1993) 787; 

C.F.Baillie and D.A.Johnston, 
Phys.Lett. B286 (1992) 44.

\bibitem{fractal1} 
H.Kawai and M.Ninomiya, 
Nucl.Phys. B336 (1990) 115;

M.E.Agishtein and A.A.Migdal, 
Nucl.Phys. B350 (1991) 690; 

N.Kawamoto, V.Kazakov, Y.Saeki and Watabiki, 
Phys.Rev.Lett. 68 (1992) 2113.

\bibitem{fractal2} 
N.Tsuda and T.Yukawa, 
Phys.Lett. B305 (1993) 223;

H.Kawai, N.Kawamoto, T.Mogami, and Y.Watabiki, 
Phys.Lett. B306 (1993) 19.

\bibitem{fishnet_analog} 
B.Sakita and M.A.Virasoro, 
Phys.Rev.Lett. 24 (1970) 1146;

H.B.Nielsen and P.Olesen, 
Phys.Lett. B32 (1970) 203.

\bibitem{circuit} 
See for example J.D.Bjorken and S.D.Drell, 
{\em McGraw Hill, New York, 1965}.

\bibitem{EDM} 
{\em Encyclopedic Dictionary of Mathematics 1980}, 
eds. S.Iyanaga and Y.Kawada, The MIT Press Cambridge, Massachusetts, 
and London, England. 

\bibitem{Wolff}
U.Wolff, Phys.Rev.Lett. 62 (1989) 361

\end{thebibliography}
\end{document}